\documentclass[aps,prb,twocolumn,showpacs,amsmath,amssymb,amsfonts,floats,floatfix,balancelastpage,dvips,superscriptaddress,reprint]{revtex4}

\usepackage{subfigure}
\usepackage{epsfig} 
\usepackage{psfrag}
\usepackage{braket}
\usepackage{color}
\usepackage{graphicx}

\usepackage{hyperref}
\hypersetup{dvips,
  pdfauthor={Christopher N. Varney, Kai Sun, Marcos Rigol, and Victor Galitski}, 
  pdftitle={Topological phase transitions for interacting finite systems}, 
  letterpaper,
  colorlinks=true,
  linkcolor=blue,
  citecolor=blue,
  pdfpagemode=UseNone,
}
\usepackage{breakurl}
\newcommand{\bbraket}[1]{\braket{\hspace{-2pt}\braket{#1}\hspace{-2pt}}}

\begin{document}

\title{Topological phase transitions for interacting finite systems}

\author{Christopher N. Varney}
\affiliation{Department of Physics, Georgetown University, Washington,
  DC 20057, USA}  
\affiliation{Joint Quantum Institute and Department of Physics, University of
  Maryland, College Park, Maryland 20742, USA}

\author{Kai Sun} 
\affiliation{Joint Quantum Institute and Department of Physics, University of
  Maryland, College Park, Maryland 20742, USA}
\affiliation{Condensed Matter Theory Center, Department of Physics, University
  of Maryland, College Park, Maryland 20742, USA}

\author{Marcos Rigol}
\affiliation{Department of Physics, Georgetown University, Washington,
  DC 20057, USA}

\author{Victor Galitski}
\affiliation{Joint Quantum Institute and Department of Physics, University of
  Maryland, College Park, Maryland 20742, USA}
\affiliation{Condensed Matter Theory Center, Department of Physics, University
  of Maryland, College Park, Maryland 20742, USA}

\begin{abstract}
  In this paper, we investigate signatures of topological phase
  transitions in interacting systems. We show that the key signature
  is the existence of a topologically protected level crossing, which
  is robust and sharply defines the topological transition, even in
  finite-size systems. 
  Spatial symmetries are argued to play a fundamental role in the
  selection of the boundary conditions to be used to locate
  topological transitions in finite systems.
  We discuss the theoretical implications of this result, and utilize
  exact diagonalization to demonstrate its manifestations in the
  Haldane-Fermi-Hubbard model. Our findings provide an efficient way
  to detect topological transitions in experiments and in numerical
  calculations that cannot access the ground-state wave function.
\end{abstract}


\maketitle


The discovery of the integer quantum Hall effect~\cite{klitzing1980}
provided a new type of quantum phase transition, the topological
transition, which does not depend on spontaneous symmetry breaking. In
recent years, the interest in topological states of matter and
topological transitions was renewed by the discovery of a class of
topological insulators with time-reversal symmetry~\cite{hasan2010,
  qiRMP}. For noninteracting insulators, it is believed that all
possible topologically ordered states have been
classified~\cite{kitaev2009,schnyder2008}, and the transition between
topologically distinct states must feature a closing of the
single-particle gap~\cite{hasan2010,qiRMP,haldane1988}. At present,
much of the recent development has focused on the role of interactions
in topological insulators, specifically the search for
interaction-induced topological
insulators~\cite{raghu2008,sun2009,uebelacker2011, dzero2010,JWen2010}
and understanding the nature of quantum phase transitions between
different topological classes~\cite{rachel2010,
  varney2010,jiang2010,hohenadler2011,zheng2011,yu2011,LWang2010,WWu2011,
  JWen2011,griset2011}.

Some of the most important challenges in the study of interacting
topological insulators lie in developing a classification scheme and
in the difficulty to accurately compute the corresponding topological
indices in an interacting system. For a Chern insulator, the standard
way of calculating the topological invariant (the Chern number)
involves determining the ground-state wave function with twisted
boundary conditions~\cite{niu1985,fukui2005} or by taking a
three-dimensional (3D) integral of the Green's
function~\cite{volovik2010, ZWang2010,gurarie2011}. Alternatively, it
has been proposed that topological order can be ascertained by
examining the single-particle gap near the edge
states~\cite{varney2010} or the entanglement spectrum~\cite{hui2008,
  fidkowski2010a}. Unfortunately, these approaches are computationally
very challenging.

In this Rapid Communication, we show that, for all interacting
topological insulators that can be classified by the Chern number, the
topological transition is sharply defined in finite-size systems. In
contrast to non-interacting infinite systems, where the topological
transition is always marked by the closing of the single-particle gap,
we show that, for an interacting finite-size system, the
interaction-driven topological transition may be characterized by the
closing of the excitation gap without necessarily closing the
single-particle gap.  This closing of the excitation gap can be viewed
as a topologically protected level crossing. Next, we show that for
models with inversion symmetry the level crossing can only occur for
boundary conditions that are invariant under inversion. Based on this
finding, we provide a prescription for efficiently determining the
topological transition and, using exact diagonalization, perform a
representative study of interacting topological states in the simplest
case of a lattice quantum Hall state with broken time-reversal
symmetry.

We begin by formulating a general theory that captures such
transitions. In contrast to ordinary quantum phase
transitions~\cite{sachdev1999}, a topological phase transition does
not require spontaneous symmetry breaking and can be precisely defined
and observed even in finite-size systems. Consider a two-dimensional
insulator with Hamiltonian $H(\lambda)$, where $\lambda$ is a control
parameter. Given twisted-boundary conditions~\cite{poilblanc1991,
  SupMat}, the Chern number can be defined~\cite{niu1985} as
\begin{align}
  C = \int \frac{d\phi_x d\phi_y}{2 \pi i} \left( \braket{\partial_{\phi_x}
      \Psi^\ast | \partial_{\phi_y} \Psi} - \braket{\partial_{\Phi_y}
      \Psi^\ast | \partial_{\phi_x} \Psi} \right),
  \label{eq:Chern_number}
\end{align}
where $\ket{\Psi}$ is the exact many-particle wave function and
$\phi_x$ ($\phi_y$) are twists along the $x$ ($y$) direction. As long
as a unique ground state is found for all twisted boundary conditions,
the integral of Eq.~\eqref{eq:Chern_number} is quantized to an integer
value for any size system. In other words, regardless of the size of the
system, we can always define a topological transition between
insulators as the place where $C$ changes its value from one integer
to another.

Now we adiabatically vary $\lambda$ from $\lambda_1$ to
$\lambda_2$. If the excitation gap remains finite during this
procedure for all twisted-boundary conditions, the value of the Chern
number must remain invariant because the topological index is
quantized to integer numbers for gapped systems. This observation
immediately implies that if the topological index changes its value,
then the excitation gap $\Delta_{\rm ex}^{(1)} = E_1 - E_0$, with
$E_0$ ($E_1$) the energy of the ground (first-excited) state, must
vanish for some twisted-boundary condition at the topological
transition. In direct contrast to an ordinary quantum phase
transition, where finite-size effects in general result in a
finite-size gap, this phenomenon of a vanishing excitation gap remains
even in finite-size systems, where the vanishing of $\Delta_{\rm
  ex}^{(1)}$ implies the existence of a level crossing between
the lowest two states. We emphasize here that this level crossing is
required by the topological properties of the ground-state wave
functions, and thus we refer it to as a topologically protected level
crossing.

Identifying this topologically protected level crossing point, in
principle, requires the computation of excitation gaps for every
twisted-boundary condition. This difficulty can be avoided if we focus
on a special class of topological transitions where (a) the system has
space-inversion symmetry and (b) the topological index changes by an
odd number at the transition.  These two conditions are satisfied in a
large class of topological transitions [including the
Haldane-Fermi-Hubbard (HFH) Hamiltonian~\cite{varney2010}, which we
investigate below as a test model]. With space-inversion symmetry, the
excitation gap at $(\phi_x,\phi_y)$ must coincide with its partner
$(-\phi_x,-\phi_y)$. At the topological transition point, this
symmetry relation implies that if the gap closes at some
$(\phi_x,\phi_y)$, then $(-\phi_x,-\phi_y)$ also has a level
crossing. Consequently, there are in general an even number of level
crossings and the Chern number must change by an even integer. In
order for the Chern number to change by an odd value at the transition
point, the level crossing must occur at one of the boundary conditions
which are their own space-inversion partners: $(0,0)$, $(\pi,0)$,
$(0,\pi)$, and $(\pi,\pi)$. Thus, we only need to examine the
excitation gap for these high-symmetry boundary conditions to identify
the topological transition. In addition, systems with higher
rotational symmetry can simplify this further~\cite{SupMat}.

Finally, we emphasize that the existence of a level crossing is a
necessary condition for a topological transition, instead of a
sufficient one. By naively looking at the excitation gap, one cannot
distinguish the topological phase transition from an accidental level
crossing. However, if one knows that the topological index does
change, e.g., by calculating the Chern number or by the use of
limiting arguments, then the level crossing must be associated with
the topological transition.

To demonstrate the physics described above, we use a thick-restart
Lanczos algorithm \cite{KWu2000} to study a model which has a lattice
quantum Hall state with broken time-reversal
symmetry~\cite{haldane1988}. Here, we consider the HFH
Hamiltonian~\cite{varney2010},
\begin{align}
  \begin{aligned}
    H = &-t_1\sum_{\braket{i \, j}} \left( c^\dag_i c^{\phantom
        \dag}_j + \text{H.c.} \right)\\ 
    &- t_2\sum_{\bbraket{i \, j}} \left(e^{i \phi_{ij}} c^\dag_i
      c^{\phantom \dag}_j + \text{H.c.} \right) + V \sum_{\braket{i \,
        j}} n_i^{\phantom \dag} n_j^{\phantom \dag} ,
  \end{aligned}
\end{align}
on a honeycomb lattice at half-filling, where $c_i^{\dag}$
($c^{\phantom\dag}_i$) are the fermion creation (annihilation)
operators at site $i$ and $n_i = c^\dag_i c_i^{\phantom\dag}$ is the
corresponding number operator. Here $t_1$ ($t_2$) are the
nearest-neighbor (next-nearest-neighbor) hopping amplitudes, $V$ is a
repulsive nearest-neighbor interaction, and the next-nearest-neighbor
hopping term has a complex phase $\phi_{ij} = \pm \phi$ corresponding
to loops in the anti-clockwise (clockwise) direction. In what follows,
we restrict our study to clusters whose symmetry in momentum space
contains the zone corner ${\bf k} = K$, as justified in
Ref.~\onlinecite{varney2010}. Also note that we set the unit of energy
$t_1 = 1$ and the definitions of all the observables are presented in
the Supplemental Material~\cite{SupMat}.

For small $V$ the system is a gapped topological insulator, which has
a unique ground state and a finite gap $\Delta_{\rm ex}^{(1)}$. In
the limit $V \rightarrow \infty$, the system turns into a
topologically trivial charge-density-wave (CDW) insulator in which all
of the particles are located on one sublattice. Here, the system
has a doubly degenerate ground state, i.e., $\Delta_{\rm ex}^{(1)} =
0$, and two finite gaps: the excitation gap $\Delta_{\rm ex}^{(2)} =
E_2 - E_0$, where $E_2$ is the energy of the second excited state, and
single-particle gap $\Delta_{\rm sp}$, which is the energy required to
add or remove a particle from the system~\cite{SupMat}. In general,
the onset of CDW order and the change in the topological index may
occur at different interaction strengths, $V_C$ and $V_T$,
respectively, opening up a topological Mott-insulating region. For
this Hamiltonian, we observe two cases that are related to the
symmetry of the cluster: (1) $V_C < V_T$ and (2) $V_C = V_T$.

\begin{figure}[tb]
  \centering
  \includegraphics[width=\columnwidth,angle=0]{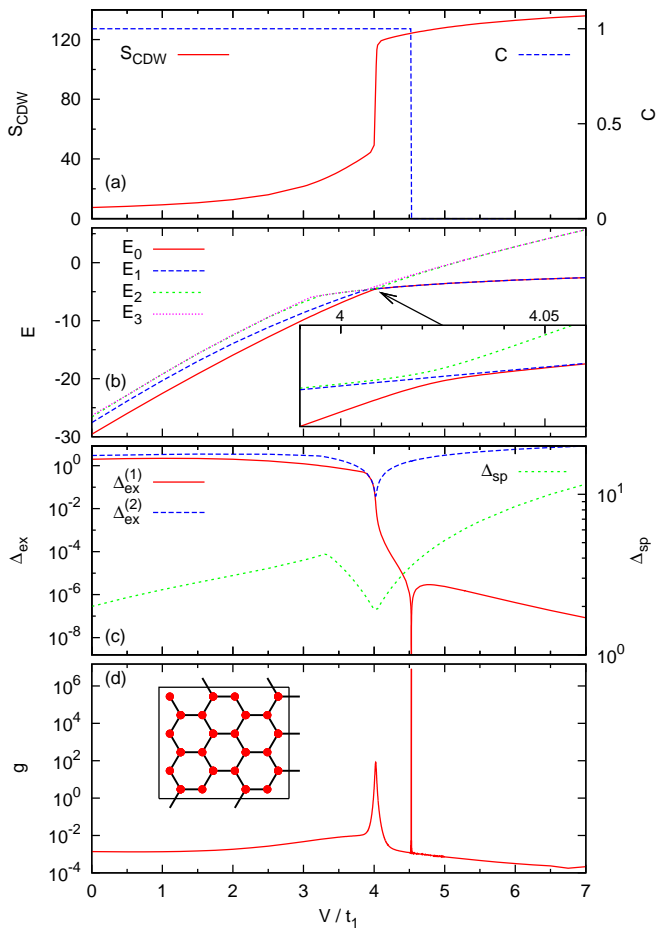}
  \caption{
    (Color online) (a) CDW structure factor $S_{\rm CDW}$ and Chern
    number $C$, (b) ground-state energy $E_0$ and first three excited
    state energies ($E_1$, $E_2$, and $E_3$) with an inset that shows
    a close up view of the avoided level crossing at $V = V_C$, (c)
    excitation gaps $\Delta_{\rm ex}^{(1,2)}$ and a single-particle
    gap $\Delta_{\rm sp}$, and (d) fidelity metric $g(V,\delta V)$
    with $\delta V = 10^{-4}$ as a function of the interaction
    strength $V / t_1$ for the $24C$ cluster [see the inset of (d)],
    and parameters $t_1 = 1.0$, $t_2 = 0.8$, and $\phi = \pi /
    2$. 
      Note that this cluster does not
      possess all of the symmetries present in the infinite system.
    \label{fig:24C}
  }
\end{figure}

Figure~\ref{fig:24C} depicts the properties of the HFH Hamiltonian for
the $24C$ cluster [see the inset in Fig.~\ref{fig:24C}(d)] with
parameters that typify the case $V_C < V_T$. In Fig.~\ref{fig:24C}(a),
we show the CDW structure factor~\cite{SupMat} and the Chern
number. Here the jump in the structure factor marks the CDW transition
at $V_C = 4.022 \pm 0.001$. In addition to the CDW transition, the
topological index also changes its value as $V$ increases, and we
identify the topological transition at $V_T = 4.5281 \pm 0.0001$.

\begin{figure}[t]
  \centering
  \includegraphics[width=\columnwidth,angle=0]{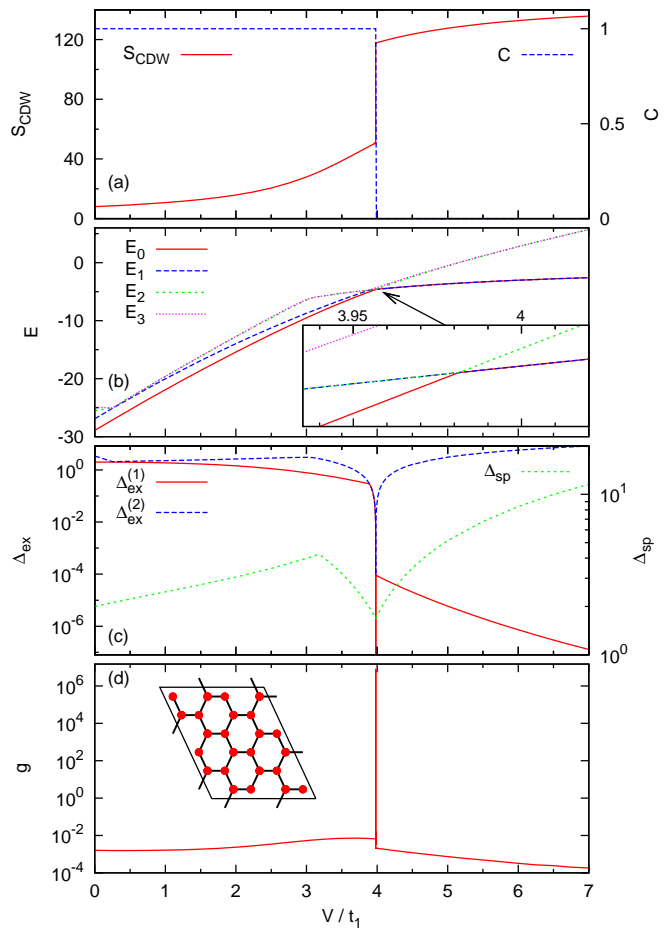}
  \caption{
    (Color online) (a) CDW structure factor $S_{\rm CDW}$ and Chern
    number $C$, (b) ground-state energy $E_0$ and first three excited
    state energies ($E_1$, $E_2$, and $E_3$) with an inset that shows
    a close up view of the level crossing at $V = V_C = V_T$, (c)
    excitation gaps $\Delta_{\rm ex}^{(1,2)}$ and a single-particle
    gap $\Delta_{\rm sp}$, and (d) fidelity metric as a function of
    the interaction strength $V / t_1$ for the $24D$ cluster [see the
    inset of (d)] and the same parameters as Fig.~\ref{fig:24C}. 
    \label{fig:24D}
  }
\end{figure}

Next, we show the four lowest-energy states in Fig.~\ref{fig:24C}(b),
with an inset focusing on the avoided level crossing at $V = V_C$. In
addition, there is a topologically protected level crossing at $V =
V_T$ (not visible). This can be seen more clearly in
Fig.~\ref{fig:24C}(c), where we show the single-particle gap
$\Delta_{\rm sp}$ and excitation gaps $\Delta_{\rm ex}^{(1)}$ and
$\Delta_{\rm ex}^{(2)}$. Here $\Delta_{\rm sp}$ and $\Delta_{\rm
  ex}^{(2)}$ both have a pronounced minimum at $V = V_C$, where both
gaps are expected to vanish in the thermodynamic limit. The
topological transition, on the other hand, is characterized by a
vanishing excitation gap $\Delta_{\rm ex}^{(1)}$, not necessarily by
the vanishing of the single-particle gap. This is in direct contrast
to Ref.~\onlinecite{LWang2010}, which claims that the topological
transition is connected to the minimum of the single-particle gap. In
addition, we emphasize that for all of the clusters we studied, the
closing of the excitation gap at the topological transition always
takes place for the periodic boundary condition case $\phi_x = \phi_y
= 0$ (Ref.~\onlinecite{SupMat}).

One natural consequence of the level crossing seen in
Fig.~\ref{fig:24C}(c) can be observed in the fidelity
metric~\cite{SupMat} $g(V,\delta V)$, which has been shown to be a
sensitive indicator of quantum phase transitions
\cite{zanardi2006,campos2007,rigol2009,varney2010,varney2011}. We
illustrate this quantity in Fig.~\ref{fig:24C}(d). While the CDW
transition is marked by a peak with finite width (independent of
$\delta V$) indicative of a traditional (first-order) phase
transition, the topological transition is characterized by a singular
point where the overlap goes to zero, and the fidelity metric has a
singular peak with height $2 / N (\delta V)^2$ and width $\sim \delta
V$, where $N$ is the number of sites. Here we emphasize that because
there is a topologically protected level crossing, this singular
behavior of the fidelity metric will always occur for a topological
transition. Due to the singular nature of this peak, numerical
calculations of $g$ may easily miss this feature, instead observing a
jump discontinuity~\cite{varney2010}.

The second case, $V_C = V_T$, is more representative of the model in
the infinite limit.  
Indeed, for any cluster which preserves the full symmetry of the
honeycomb lattice we find that $V_C = V_T$ for all parameters studied.
In Fig.~\ref{fig:24D}, we show the properties of the $24D$ cluster
[see the inset in Fig.~\ref{fig:24D}(d)] with the same parameters as
in Fig.~\ref{fig:24C}. In Fig.~\ref{fig:24D}(a), the CDW transition,
marked by the jump in the structure factor, and the change in the
Chern number both occur at $V = V_C = V_T = 3.9813 \pm 0.0001$. As a
result, there is a level crossing [shown in Fig.~\ref{fig:24D}(b)]
which is topologically protected and the ground state is triply
degenerate. This three-fold degeneracy becomes evident by examining
the excitation gaps $\Delta_{\rm ex}^{(1)}$ and $\Delta_{\rm
  ex}^{(2)}$ [Fig.~\ref{fig:24D}(c)], which both approach zero as $V
\to V_T$. In the fidelity metric [Fig.~\ref{fig:24D}(d)], we see only
the singular peak associated with the topological transition. In
general, we find that the two features do coexist, with the CDW
fidelity peak becoming much sharper and, in many cases, completely
obscured by the singular peak at the topological transition.

\begin{figure}[t]
  \centering
  \includegraphics[height=\columnwidth,angle=-90]{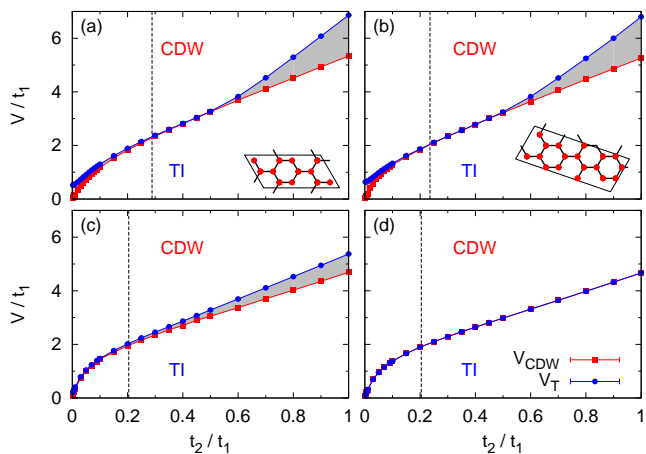}
  \caption{ 
    (Color online) $t_2$-$V$ phase diagram for the (a) $12A$, (b)
    $18C$, (c) $24C$, and (d) $24D$ clusters with parameters $t_1 =
    1.0$ and $\phi = \pi / 2$. The red squares indicate the onset of
    CDW order, and the blue circles indicate the topological
    transition. The shaded region marks the coexistence of CDW and
    topological order. All points to the vertical line are dominated
    by strong finite-size effects. Illustrations of the $12A$ and
    $18C$ clusters are shown as insets in the corresponding panel.
    \label{fig:pd}
  }
\end{figure}

In Fig.~\ref{fig:pd}, we show the $t_2$-$V$ phase diagrams for
clusters from 12 to 24 sites. In general, the system is a topological
insulator (TI) at weak-coupling and a topologically trivial CDW
insulator at strong-coupling. For clusters without six-fold rotational
symmetry [Figs.~\ref{fig:pd}(a)-\ref{fig:pd}(c)], we find a region of
parameter space with a coexistence of topological and CDW order. For
clusters with six-fold rotational symmetry [Fig.~\ref{fig:pd}(d)], we
find that the CDW and topological transitions always coincide. As this
symmetry is present in the thermodynamic limit and given the reduction
we see in the region where CDW and topological order coexist as we
increase the size of the clusters without that symmetry, we conclude
that the topological CDW insulator phase does not exist in this model
in the thermodynamic limit. In addition, we find no evidence for a
topologically trivial insulating phase without CDW
order~\cite{LWang2010}.

It is known that, at $t_2 = 0$, the system exhibits a semi-metal-CDW
transition at finite $V$ in the thermodynamic limit. However, in
Fig.~\ref{fig:pd}, we observe a sudden decrease in $V_C$ for small
$t_2$, resulting in a very small value of $V_C$ at $t_2=0$. This
apparent contradiction is due to finite-size effects, which become
dominant as $V,t_2 \to 0$. In this region, the single-particle gap
becomes very small ($\sim t_2$), so much larger systems (with sizes
larger than the inverse gap) need to be studied to accurately
determine the phase boundaries. To highlight this observation, we mark
the onset of strong finite-size effects ($t_2 =L^{-1}$) by a dashed
line in Fig.~\ref{fig:pd}, where $L$ is the linear size of the system.
The drop of $V_C$ takes place in the regime with strong finite-size
effects ($t_2 < L^{-1}$). As the system size increases, the region
with $t_2 < L^{-1}$ is pushed to $t_2 = 0$, so the sudden drop of
$V_C$ at small $t_2$ disappears in the thermodynamic limit, resulting
in a finite $V_C$ in the limit $t_2 \to 0$.

In summary, we have shown that the closing of the excitation gap is a
signature of the topological transition in interacting systems. This
topologically protected level crossing exists even in finite-size
systems, and the resulting singular behavior at the transition can be
observed in quantities such as the fidelity metric.
When coupled with the use of spatial symmetries to simplify the choice
of boundary conditions, this phenomenon provides an efficient scheme
for locating the topological transition which
can be straightforwardly generalized to time-reversal invariant
topological insulators and fractional topological states. Aside from a
few special cases \cite{alba2011,zhao2011}, our findings provide a
generic methodology to locate a topological transition in interacting
finite-size systems via an experimentally measurable
quantity. Consequently, this allows for the study of topological
phases and transitions in cold-atom experiments and computational
approaches in which the ground-state wave function is not accessible.

  This research was supported by NSF through JQI-PFC (C.N.V. and
  K.S.), ONR (C.N.V. and M.R.), and U.S.-ARO (V.G.).


\bibliography{references}

\begin{thebibliography}{38}
\expandafter\ifx\csname natexlab\endcsname\relax\def\natexlab#1{#1}\fi
\expandafter\ifx\csname bibnamefont\endcsname\relax
  \def\bibnamefont#1{#1}\fi
\expandafter\ifx\csname bibfnamefont\endcsname\relax
  \def\bibfnamefont#1{#1}\fi
\expandafter\ifx\csname citenamefont\endcsname\relax
  \def\citenamefont#1{#1}\fi
\expandafter\ifx\csname url\endcsname\relax
  \def\url#1{\texttt{#1}}\fi
\expandafter\ifx\csname urlprefix\endcsname\relax\def\urlprefix{URL }\fi
\providecommand{\bibinfo}[2]{#2}
\providecommand{\eprint}[2][]{\url{#2}}

\bibitem[{\citenamefont{{von Klitzing} et~al.}(1980)\citenamefont{{von
  Klitzing}, Dorda, and Pepper}}]{klitzing1980}
\bibinfo{author}{\bibfnamefont{K.}~\bibnamefont{{von Klitzing}}},
  \bibinfo{author}{\bibfnamefont{G.}~\bibnamefont{Dorda}}, \bibnamefont{and}
  \bibinfo{author}{\bibfnamefont{M.}~\bibnamefont{Pepper}},
  \bibinfo{journal}{Phys. Rev. Lett.} \textbf{\bibinfo{volume}{45}},
  \bibinfo{pages}{494} (\bibinfo{year}{1980}).

\bibitem[{\citenamefont{Hasan and Kane}(2010)}]{hasan2010}
\bibinfo{author}{\bibfnamefont{M.~Z.} \bibnamefont{Hasan}} \bibnamefont{and}
  \bibinfo{author}{\bibfnamefont{C.~L.} \bibnamefont{Kane}},
  \bibinfo{journal}{Rev. Mod. Phys.} \textbf{\bibinfo{volume}{82}},
  \bibinfo{pages}{3045} (\bibinfo{year}{2010}).

\bibitem[{\citenamefont{Qi and Zhang}(2011)}]{qiRMP}
\bibinfo{author}{\bibfnamefont{X.-L.} \bibnamefont{Qi}} \bibnamefont{and}
  \bibinfo{author}{\bibfnamefont{S.-C.} \bibnamefont{Zhang}},
  \bibinfo{journal}{Rev. Mod. Phys.} \textbf{\bibinfo{volume}{83}},
  \bibinfo{pages}{1057} (\bibinfo{year}{2011}).

\bibitem[{\citenamefont{Kitaev}(2009)}]{kitaev2009}
\bibinfo{author}{\bibfnamefont{A.}~\bibnamefont{Kitaev}}, in
  \emph{\bibinfo{booktitle}{Advances in Theoretical Physics: Landau Memorial
  Conference}}, edited by
  \bibinfo{editor}{\bibfnamefont{V.}~\bibnamefont{Lebedev}} \bibnamefont{and}
  \bibinfo{editor}{\bibfnamefont{M.}~\bibnamefont{Feigel'man}}
  (\bibinfo{publisher}{AIP}, \bibinfo{address}{Melville, NY},
  \bibinfo{year}{2009}), vol. \bibinfo{volume}{1134} of
  \emph{\bibinfo{series}{AIP Conf. Proc.}}, pp. \bibinfo{pages}{22--30}.

\bibitem[{\citenamefont{Schnyder et~al.}(2008)\citenamefont{Schnyder, Ryu,
  Furusaki, and Ludwig}}]{schnyder2008}
\bibinfo{author}{\bibfnamefont{A.~P.} \bibnamefont{Schnyder}},
  \bibinfo{author}{\bibfnamefont{S.}~\bibnamefont{Ryu}},
  \bibinfo{author}{\bibfnamefont{A.}~\bibnamefont{Furusaki}}, \bibnamefont{and}
  \bibinfo{author}{\bibfnamefont{A.~W.~W.} \bibnamefont{Ludwig}},
  \bibinfo{journal}{Phys. Rev. B} \textbf{\bibinfo{volume}{78}},
  \bibinfo{pages}{195125} (\bibinfo{year}{2008}).

\bibitem[{\citenamefont{Haldane}(1988)}]{haldane1988}
\bibinfo{author}{\bibfnamefont{F.~D.~M.} \bibnamefont{Haldane}},
  \bibinfo{journal}{Phys. Rev. Lett.} \textbf{\bibinfo{volume}{61}},
  \bibinfo{pages}{2015} (\bibinfo{year}{1988}).

\bibitem[{\citenamefont{Raghu et~al.}(2008)\citenamefont{Raghu, Qi, Honerkamp,
  and Zhang}}]{raghu2008}
\bibinfo{author}{\bibfnamefont{S.}~\bibnamefont{Raghu}},
  \bibinfo{author}{\bibfnamefont{X.-L.} \bibnamefont{Qi}},
  \bibinfo{author}{\bibfnamefont{C.}~\bibnamefont{Honerkamp}},
  \bibnamefont{and} \bibinfo{author}{\bibfnamefont{S.-C.} \bibnamefont{Zhang}},
  \bibinfo{journal}{Phys. Rev. Lett.} \textbf{\bibinfo{volume}{100}},
  \bibinfo{pages}{156401} (\bibinfo{year}{2008}).

\bibitem[{\citenamefont{Sun et~al.}(2009)\citenamefont{Sun, Yao, Fradkin, and
  Kivelson}}]{sun2009}
\bibinfo{author}{\bibfnamefont{K.}~\bibnamefont{Sun}},
  \bibinfo{author}{\bibfnamefont{H.}~\bibnamefont{Yao}},
  \bibinfo{author}{\bibfnamefont{E.}~\bibnamefont{Fradkin}}, \bibnamefont{and}
  \bibinfo{author}{\bibfnamefont{S.~A.} \bibnamefont{Kivelson}},
  \bibinfo{journal}{Phys. Rev. Lett.} \textbf{\bibinfo{volume}{103}},
  \bibinfo{pages}{046811} (\bibinfo{year}{2009}).

\bibitem[{\citenamefont{Uebelacker and Honerkamp}(2011)}]{uebelacker2011}
\bibinfo{author}{\bibfnamefont{S.}~\bibnamefont{Uebelacker}} \bibnamefont{and}
  \bibinfo{author}{\bibfnamefont{C.}~\bibnamefont{Honerkamp}},
  \bibinfo{journal}{Phys. Rev. B} \textbf{\bibinfo{volume}{84}},
  \bibinfo{pages}{205122} (\bibinfo{year}{2011}).

\bibitem[{\citenamefont{Dzero et~al.}(2010)\citenamefont{Dzero, Sun, Galitski,
  and Coleman}}]{dzero2010}
\bibinfo{author}{\bibfnamefont{M.}~\bibnamefont{Dzero}},
  \bibinfo{author}{\bibfnamefont{K.}~\bibnamefont{Sun}},
  \bibinfo{author}{\bibfnamefont{V.}~\bibnamefont{Galitski}}, \bibnamefont{and}
  \bibinfo{author}{\bibfnamefont{P.}~\bibnamefont{Coleman}},
  \bibinfo{journal}{Phys. Rev. Lett.} \textbf{\bibinfo{volume}{104}},
  \bibinfo{pages}{106408} (\bibinfo{year}{2010}).

\bibitem[{\citenamefont{Wen et~al.}(2010)\citenamefont{Wen, R\"uegg,
  Joseph~Wang, and Fiete}}]{JWen2010}
\bibinfo{author}{\bibfnamefont{J.}~\bibnamefont{Wen}},
  \bibinfo{author}{\bibfnamefont{A.}~\bibnamefont{R\"uegg}},
  \bibinfo{author}{\bibfnamefont{C.-C.} \bibnamefont{Joseph~Wang}},
  \bibnamefont{and} \bibinfo{author}{\bibfnamefont{G.~A.} \bibnamefont{Fiete}},
  \bibinfo{journal}{Phys. Rev. B} \textbf{\bibinfo{volume}{82}},
  \bibinfo{pages}{075125} (\bibinfo{year}{2010}).

\bibitem[{\citenamefont{Rachel and Le~Hur}(2010)}]{rachel2010}
\bibinfo{author}{\bibfnamefont{S.}~\bibnamefont{Rachel}} \bibnamefont{and}
  \bibinfo{author}{\bibfnamefont{K.}~\bibnamefont{Le~Hur}},
  \bibinfo{journal}{Phys. Rev. B} \textbf{\bibinfo{volume}{82}},
  \bibinfo{pages}{075106} (\bibinfo{year}{2010}).

\bibitem[{\citenamefont{Varney et~al.}(2010)\citenamefont{Varney, Sun, Rigol,
  and Galitski}}]{varney2010}
\bibinfo{author}{\bibfnamefont{C.~N.} \bibnamefont{Varney}},
  \bibinfo{author}{\bibfnamefont{K.}~\bibnamefont{Sun}},
  \bibinfo{author}{\bibfnamefont{M.}~\bibnamefont{Rigol}}, \bibnamefont{and}
  \bibinfo{author}{\bibfnamefont{V.}~\bibnamefont{Galitski}},
  \bibinfo{journal}{Phys. Rev. B} \textbf{\bibinfo{volume}{82}},
  \bibinfo{pages}{115125} (\bibinfo{year}{2010}).

\bibitem[{\citenamefont{Jiang et~al.}(2010)\citenamefont{Jiang, Rachel, Weng,
  Zhang, and Wang}}]{jiang2010}
\bibinfo{author}{\bibfnamefont{H.-C.} \bibnamefont{Jiang}},
  \bibinfo{author}{\bibfnamefont{S.}~\bibnamefont{Rachel}},
  \bibinfo{author}{\bibfnamefont{Z.-Y.} \bibnamefont{Weng}},
  \bibinfo{author}{\bibfnamefont{S.-C.} \bibnamefont{Zhang}}, \bibnamefont{and}
  \bibinfo{author}{\bibfnamefont{Z.}~\bibnamefont{Wang}},
  \bibinfo{journal}{Phys. Rev. B} \textbf{\bibinfo{volume}{82}},
  \bibinfo{pages}{220403} (\bibinfo{year}{2010}).

\bibitem[{\citenamefont{Hohenadler et~al.}(2011)\citenamefont{Hohenadler, Lang,
  and Assaad}}]{hohenadler2011}
\bibinfo{author}{\bibfnamefont{M.}~\bibnamefont{Hohenadler}},
  \bibinfo{author}{\bibfnamefont{T.~C.} \bibnamefont{Lang}}, \bibnamefont{and}
  \bibinfo{author}{\bibfnamefont{F.~F.} \bibnamefont{Assaad}},
  \bibinfo{journal}{Phys. Rev. Lett.} \textbf{\bibinfo{volume}{106}},
  \bibinfo{pages}{100403} (\bibinfo{year}{2011}).

\bibitem[{\citenamefont{Zheng et~al.}(2011)\citenamefont{Zheng, Zhang, and
  Wu}}]{zheng2011}
\bibinfo{author}{\bibfnamefont{D.}~\bibnamefont{Zheng}},
  \bibinfo{author}{\bibfnamefont{G.-M.} \bibnamefont{Zhang}}, \bibnamefont{and}
  \bibinfo{author}{\bibfnamefont{C.}~\bibnamefont{Wu}}, \bibinfo{journal}{Phys.
  Rev. B} \textbf{\bibinfo{volume}{84}}, \bibinfo{pages}{205121}
  (\bibinfo{year}{2011}).

\bibitem[{\citenamefont{Yu et~al.}(2011)\citenamefont{Yu, Xie, and
  Li}}]{yu2011}
\bibinfo{author}{\bibfnamefont{S.-L.} \bibnamefont{Yu}},
  \bibinfo{author}{\bibfnamefont{X.~C.} \bibnamefont{Xie}}, \bibnamefont{and}
  \bibinfo{author}{\bibfnamefont{J.-X.} \bibnamefont{Li}},
  \bibinfo{journal}{Phys. Rev. Lett.} \textbf{\bibinfo{volume}{107}},
  \bibinfo{pages}{010401} (\bibinfo{year}{2011}).

\bibitem[{\citenamefont{Wang et~al.}()\citenamefont{Wang, Shi, Zhang, Wang,
  Dai, and Xie}}]{LWang2010}
\bibinfo{author}{\bibfnamefont{L.}~\bibnamefont{Wang}},
  \bibinfo{author}{\bibfnamefont{H.}~\bibnamefont{Shi}},
  \bibinfo{author}{\bibfnamefont{S.}~\bibnamefont{Zhang}},
  \bibinfo{author}{\bibfnamefont{X.}~\bibnamefont{Wang}},
  \bibinfo{author}{\bibfnamefont{X.}~\bibnamefont{Dai}}, \bibnamefont{and}
  \bibinfo{author}{\bibfnamefont{X.~C.} \bibnamefont{Xie}},
  \bibinfo{note}{{arXiv}:1012.5163}.

\bibitem[{\citenamefont{Wu et~al.}()\citenamefont{Wu, Rachel, Liu, and
  Hur}}]{WWu2011}
\bibinfo{author}{\bibfnamefont{W.}~\bibnamefont{Wu}},
  \bibinfo{author}{\bibfnamefont{S.}~\bibnamefont{Rachel}},
  \bibinfo{author}{\bibfnamefont{W.-M.} \bibnamefont{Liu}}, \bibnamefont{and}
  \bibinfo{author}{\bibfnamefont{K.~L.} \bibnamefont{Hur}},
  \bibinfo{note}{{arXiv}:1106.0943}.

\bibitem[{\citenamefont{Wen et~al.}()\citenamefont{Wen, Kargarian, Vaezi, and
  Fiete}}]{JWen2011}
\bibinfo{author}{\bibfnamefont{J.}~\bibnamefont{Wen}},
  \bibinfo{author}{\bibfnamefont{M.}~\bibnamefont{Kargarian}},
  \bibinfo{author}{\bibfnamefont{A.}~\bibnamefont{Vaezi}}, \bibnamefont{and}
  \bibinfo{author}{\bibfnamefont{G.~A.} \bibnamefont{Fiete}},
  \bibinfo{note}{{arXiv}:1107.0007 (unpublished)}.

\bibitem[{\citenamefont{Griset and Xu}()}]{griset2011}
\bibinfo{author}{\bibfnamefont{C.}~\bibnamefont{Griset}} \bibnamefont{and}
  \bibinfo{author}{\bibfnamefont{C.}~\bibnamefont{Xu}},
  \bibinfo{note}{{arXiv}:1107.1245}.

\bibitem[{\citenamefont{Niu et~al.}(1985)\citenamefont{Niu, Thouless, and
  Wu}}]{niu1985}
\bibinfo{author}{\bibfnamefont{Q.}~\bibnamefont{Niu}},
  \bibinfo{author}{\bibfnamefont{D.~J.} \bibnamefont{Thouless}},
  \bibnamefont{and} \bibinfo{author}{\bibfnamefont{Y.-S.} \bibnamefont{Wu}},
  \bibinfo{journal}{Phys. Rev. B} \textbf{\bibinfo{volume}{31}},
  \bibinfo{pages}{3372} (\bibinfo{year}{1985}).

\bibitem[{\citenamefont{Fukui et~al.}(2005)\citenamefont{Fukui, Hatsugai, and
  Suzuki}}]{fukui2005}
\bibinfo{author}{\bibfnamefont{T.}~\bibnamefont{Fukui}},
  \bibinfo{author}{\bibfnamefont{Y.}~\bibnamefont{Hatsugai}}, \bibnamefont{and}
  \bibinfo{author}{\bibfnamefont{H.}~\bibnamefont{Suzuki}},
  \bibinfo{journal}{J. Phys. Soc. Jpn.} \textbf{\bibinfo{volume}{74}},
  \bibinfo{pages}{1674} (\bibinfo{year}{2005}).

\bibitem[{\citenamefont{Volovik}(2010)}]{volovik2010}
\bibinfo{author}{\bibfnamefont{G.~E.} \bibnamefont{Volovik}},
  \bibinfo{journal}{JETP Lett.} \textbf{\bibinfo{volume}{91}},
  \bibinfo{pages}{55} (\bibinfo{year}{2010}).

\bibitem[{\citenamefont{Wang et~al.}(2010)\citenamefont{Wang, Qi, and
  Zhang}}]{ZWang2010}
\bibinfo{author}{\bibfnamefont{Z.}~\bibnamefont{Wang}},
  \bibinfo{author}{\bibfnamefont{X.-L.} \bibnamefont{Qi}}, \bibnamefont{and}
  \bibinfo{author}{\bibfnamefont{S.-C.} \bibnamefont{Zhang}},
  \bibinfo{journal}{Phys. Rev. Lett.} \textbf{\bibinfo{volume}{105}},
  \bibinfo{pages}{256803} (\bibinfo{year}{2010}).

\bibitem[{\citenamefont{Gurarie}(2011)}]{gurarie2011}
\bibinfo{author}{\bibfnamefont{V.}~\bibnamefont{Gurarie}},
  \bibinfo{journal}{Phys. Rev. B} \textbf{\bibinfo{volume}{83}},
  \bibinfo{pages}{085426} (\bibinfo{year}{2011}).

\bibitem[{\citenamefont{Li and Haldane}(2008)}]{hui2008}
\bibinfo{author}{\bibfnamefont{H.}~\bibnamefont{Li}} \bibnamefont{and}
  \bibinfo{author}{\bibfnamefont{F.~D.~M.} \bibnamefont{Haldane}},
  \bibinfo{journal}{Phys. Rev. Lett.} \textbf{\bibinfo{volume}{101}},
  \bibinfo{pages}{010504} (\bibinfo{year}{2008}).

\bibitem[{\citenamefont{Fidkowski}(2010)}]{fidkowski2010a}
\bibinfo{author}{\bibfnamefont{L.}~\bibnamefont{Fidkowski}},
  \bibinfo{journal}{Phys. Rev. Lett.} \textbf{\bibinfo{volume}{104}},
  \bibinfo{pages}{130502} (\bibinfo{year}{2010}).

\bibitem[{\citenamefont{Sachdev}(1999)}]{sachdev1999}
\bibinfo{author}{\bibfnamefont{S.}~\bibnamefont{Sachdev}},
  \emph{\bibinfo{title}{Quantum Phase Transitions}}
  (\bibinfo{publisher}{Cambridge University Press},
  \bibinfo{address}{Cambridge, UK}, \bibinfo{year}{1999}).

\bibitem[{\citenamefont{Poilblanc}(1991)}]{poilblanc1991}
\bibinfo{author}{\bibfnamefont{D.}~\bibnamefont{Poilblanc}},
  \bibinfo{journal}{Phys. Rev. B} \textbf{\bibinfo{volume}{44}},
  \bibinfo{pages}{9562} (\bibinfo{year}{1991}).

\bibitem[{Sup()}]{SupMat}
\bibinfo{note}{See Supplemental Material at
  \url{http://link.aps.org/supplemental/10.1103/PhysRevB.84.241105} for
  additional theoretical discussion, a full description of the observables, and
  additional results.}

\bibitem[{\citenamefont{Wu and Simon}(2000)}]{KWu2000}
\bibinfo{author}{\bibfnamefont{K.}~\bibnamefont{Wu}} \bibnamefont{and}
  \bibinfo{author}{\bibfnamefont{H.}~\bibnamefont{Simon}},
  \bibinfo{journal}{SIAM. J. Matrix Anal. Appl.} \textbf{\bibinfo{volume}{22}},
  \bibinfo{pages}{602} (\bibinfo{year}{2000}).

\bibitem[{\citenamefont{Zanardi and Paunkovi\ifmmode~\acute{c}\else
  \'{c}\fi{}}(2006)}]{zanardi2006}
\bibinfo{author}{\bibfnamefont{P.}~\bibnamefont{Zanardi}} \bibnamefont{and}
  \bibinfo{author}{\bibfnamefont{N.}~\bibnamefont{Paunkovi\ifmmode~\acute{c}\e%
lse \'{c}\fi{}}}, \bibinfo{journal}{Phys. Rev. E}
  \textbf{\bibinfo{volume}{74}}, \bibinfo{pages}{031123}
  (\bibinfo{year}{2006}).

\bibitem[{\citenamefont{Campos~Venuti and Zanardi}(2007)}]{campos2007}
\bibinfo{author}{\bibfnamefont{L.}~\bibnamefont{Campos~Venuti}}
  \bibnamefont{and} \bibinfo{author}{\bibfnamefont{P.}~\bibnamefont{Zanardi}},
  \bibinfo{journal}{Phys. Rev. Lett.} \textbf{\bibinfo{volume}{99}},
  \bibinfo{pages}{095701} (\bibinfo{year}{2007}).

\bibitem[{\citenamefont{Rigol et~al.}(2009)\citenamefont{Rigol, Shastry, and
  Haas}}]{rigol2009}
\bibinfo{author}{\bibfnamefont{M.}~\bibnamefont{Rigol}},
  \bibinfo{author}{\bibfnamefont{B.~S.} \bibnamefont{Shastry}},
  \bibnamefont{and} \bibinfo{author}{\bibfnamefont{S.}~\bibnamefont{Haas}},
  \bibinfo{journal}{Phys. Rev. B} \textbf{\bibinfo{volume}{80}},
  \bibinfo{pages}{094529} (\bibinfo{year}{2009}).

\bibitem[{\citenamefont{Varney et~al.}(2011)\citenamefont{Varney, Sun,
  Galitski, and Rigol}}]{varney2011}
\bibinfo{author}{\bibfnamefont{C.~N.} \bibnamefont{Varney}},
  \bibinfo{author}{\bibfnamefont{K.}~\bibnamefont{Sun}},
  \bibinfo{author}{\bibfnamefont{V.}~\bibnamefont{Galitski}}, \bibnamefont{and}
  \bibinfo{author}{\bibfnamefont{M.}~\bibnamefont{Rigol}},
  \bibinfo{journal}{Phys. Rev. Lett.} \textbf{\bibinfo{volume}{107}},
  \bibinfo{pages}{077201} (\bibinfo{year}{2011}).

\bibitem[{\citenamefont{Alba et~al.}(2011)\citenamefont{Alba,
  Fernandez-Gonzalvo, Mur-Petit, Pachos, and Garcia-Ripoll}}]{alba2011}
\bibinfo{author}{\bibfnamefont{E.}~\bibnamefont{Alba}},
  \bibinfo{author}{\bibfnamefont{X.}~\bibnamefont{Fernandez-Gonzalvo}},
  \bibinfo{author}{\bibfnamefont{J.}~\bibnamefont{Mur-Petit}},
  \bibinfo{author}{\bibfnamefont{J.~K.} \bibnamefont{Pachos}},
  \bibnamefont{and} \bibinfo{author}{\bibfnamefont{J.~J.}
  \bibnamefont{Garcia-Ripoll}}, \bibinfo{journal}{Phys. Rev. Lett.}
  \textbf{\bibinfo{volume}{107}}, \bibinfo{pages}{235301}
  (\bibinfo{year}{2011}).

\bibitem[{\citenamefont{Zhao et~al.}()\citenamefont{Zhao, Bray-Ali, Williams,
  Spielman, and Satija}}]{zhao2011}
\bibinfo{author}{\bibfnamefont{E.}~\bibnamefont{Zhao}},
  \bibinfo{author}{\bibfnamefont{N.}~\bibnamefont{Bray-Ali}},
  \bibinfo{author}{\bibfnamefont{C.~J.} \bibnamefont{Williams}},
  \bibinfo{author}{\bibfnamefont{I.~B.} \bibnamefont{Spielman}},
  \bibnamefont{and} \bibinfo{author}{\bibfnamefont{I.~I.}
  \bibnamefont{Satija}}, \bibinfo{note}{{arXiv}:1105.3100}.

\end{thebibliography}

\end{document}